\documentclass[aps,prd,twocolumn,showpacs,floats,floatfix,letterpaper,nofootinbib,superscriptaddress,]{revtex4}

\usepackage{amssymb,amsmath,latexsym,mathrsfs}
\usepackage{graphicx}
\usepackage{epsfig}
\usepackage{multirow}
\usepackage{array}

\newcommand{\neff}{N_{\textrm{eff}}}

\newcommand{\mnu}{m_{\nu}}

\begin{document}


\title{Testing standard and non-standard neutrino physics with cosmological data}

\author{Elena Giusarma}
\affiliation{IFIC, Universidad de Valencia-CSIC, 46071, Valencia, Spain}
\author{Roland de Putter}
\affiliation{Jet Propulsion Laboratory, California Institute of Technology, Pasadena, CA 91109, USA}
\affiliation{California Institute of Technology, Pasadena, CA 91125, USA}
\author{Olga Mena$^1$}

\begin{abstract}
Cosmological constraints on the sum of neutrino masses and on the effective number of neutrino species in standard and non-standard
scenarios are computed using the most recent available cosmological data. Our cosmological data sets include the measurement of the
Baryonic Acoustic Oscillation (BAO) feature in the Data Release 9 CMASS sample of the Baryon Oscillation Spectroscopic Survey (BOSS).
We study in detail the different degeneracies among the parameters, as well as the impact of the different data sets used in the analyses.
When considering bounds on the sum of the three active neutrino masses, the information in the BAO signal from galaxy clustering
measurements is approximately equally powerful as the shape information from the matter power spectrum. The most stringent bound we find
is $\sum m_\nu<0.32$~eV at 95$\%$ CL. When non-standard neutrino scenarios with $\neff$ massless or massive neutrino species are examined,
power spectrum shape measurements provide slightly better bounds than the BAO signal only, due to the breaking of parameter degeneracies.

Cosmic Microwave Background data from high multipoles from the South Pole Telescope turns out to be crucial for extracting the
number of effective neutrino species. Recent BOSS data combined with CMB and Hubble Space Telescope measurements give $\neff=3.66^{+0.20 +0.73}_{-0.21 -0.69}$ in the massless neutrino scenario, and similar results are obtained in the massive case. The evidence for extra radiation $\neff>3$ often claimed in
the literature therefore remains at the $2\sigma$ level when considering up-to-date cosmological data sets.
Measurements from the Wilkinson Microwave Anisotropy Probe combined with a prior on the Hubble parameter from the Hubble Space Telescope are very powerful in constraining either the sum of the three active neutrino masses or
the number of massless neutrino species. If the former two parameters are allowed to freely vary, however, the bounds from the combination of these two cosmological probes get worse by an order of magnitude.

\end{abstract}


\maketitle

\section{Introduction}

Over decades, solar, atmospheric, reactor, and accelerator neutrinos have provided compelling evidence for the existence of neutrino oscillations, implying non-zero neutrino masses. Oscillation experiments only provide bounds on the neutrino mass squared differences, and
therefore information on the overall scale of the neutrino mass must come from other experiments. Cosmological data provides a tool to test the absolute scale of neutrino masses. Neutrino masses and abundances leave crucial features in several cosmological observables. 
The amount of primordial relativistic neutrinos affects Cosmic Microwave Background (CMB) anisotropies and non-relativistic neutrinos in the recent Universe 
suppress the growth of matter density fluctuations and galaxy clustering. 
Cosmology can therefore weigh neutrinos, providing an upper bound on the
sum of the three active neutrino masses, $\sum m_\nu \sim 0.26$~eV at $95\%$ confidence level (CL)~\cite{dePutter:2012sh} (see also Refs.\cite{Komatsu:2010fb,Reid:2009nq,Hamann:2010pw}).
The former bound is found when CMB measurements from the Wilkinson Microwave Anisotropy Probe (WMAP) are
combined with information on the distribution of galaxies based on a catalog of photometric galaxy redshifts in Sloan Digital
Sky Survey III Data Release Eight (SDSS DR8), and with the most recent measurement of the Hubble parameter from the
Hubble Space Telescope (HST). It assumes a flat universe with a cosmological constant, i.e.~a $\Lambda$CDM cosmology. 

In the Standard Model of elementary particles, there are three active neutrinos.
However, there is no fundamental symmetry in nature establishing a definite number of right-handed (sterile) neutrino species. In fact, models with additional $\sim 1$~eV massive sterile neutrinos~\cite{Sorel:2003hf} have been introduced to explain short baseline (SBL) antineutrino data~\cite{Aguilar:2001ty} in terms of neutrino oscillations. Up to date cosmological constraints on massive sterile and active neutrino species have been presented in different cosmological scenarios using different data sets, see Refs.~\cite{rt,us,Hamann,Giusarma,Joudaki,latest}. 

In addition, extra relativistic degrees of freedom could also arise from different physics, such as massless sterile neutrino species,
axions~\cite{axions}, decay of non-relativistic matter~\cite{decay}, gravity waves~\cite{gw}, extra dimensions~\cite{extra}, early dark energy~\cite{ede} or asymmetric dark matter models~\cite{Blennow:2012de}. Cosmological probes have been extensively exploited in the literature to set bounds on the relativistic energy density of the universe in terms of the effective number of neutrinos $\neff$ (see Refs.~\cite{Mangano:2006ur,Hamann:2007pi,Reid:2009nq,Komatsu:2010fb,Hamann:2010pw,Hamann:2011hu,Nollett:2011aa}). Measurements of Cosmic Microwave Background (CMB) anisotropies at arc-minute angular scales from the South Pole Telescope (SPT)~\cite{spt} and the Atacama Cosmology Telescope (ACT)~\cite{act}, when combined with other cosmological data sets have provided the constraint $\neff=4.08_{-0.68}^{+0.71}$ at $95 \%$ CL~\cite{darkr,Hou:2011ec,Smith:2011es,darkr2}), showing evidence for $\neff>0$ at more than $7\sigma$ and suggesting values higher than those expected in the canonical scenario. If the effective number of neutrino species $\neff$ is larger than the Standard Model prediction, i.e.~if $\neff>3.046$ during the Big Bang Nucleosynthesis (BBN) era, the Hubble expansion rate will be larger, causing therefore weak interactions to freeze out earlier.
The standard BBN predictions for light element abundances will change. Current analyses of the observed primordial abundances indicate best-fit values of $\Delta \neff \sim 0.5 - 0.8$~\cite{Izotov:2010ca,Mangano:2010ei,rt,Hamann}, and bounds $\Delta \neff <1-2$ at $95 \%$ CL, depending on the assumptions on the fiducial cosmology, and are consistent with the canonical $\Delta \neff = 0$.
Nevertheless, the extra radiation species do not necessarily feature thermal abundances at BBN~\cite{Melchiorri:2008gq,Acero:2008rh}
so that the true number of additional species could possibly be larger than the effective number defined in terms of thermal abundances.

Here we follow a general approach and study the constraints on several neutrino scenarios with the most recent available cosmological data.
We consider both standard neutrino scenarios with three active massive neutrinos as well as more general, non-standard schemes with $\neff$ massless
or $\neff$ massive species.
Of the two scenarios with free number of neutrino species, $N_{\rm eff}$, the latter one is the most realistic and appropriate scenario,
since we know observationally that neutrinos have mass. However, this scenario
has been explored much less in the literature than the case of $N_{\rm eff}$ neutrinos with mass fixed to zero (although see, e.g., \cite{us,Giusarma}
and the very recent \cite{latest}).
In fact, we will show that in the scenario with
$\neff$ massive neutrino species, the neutrino constraints from some combinations of cosmological data sets are much worse
than in the other two simpler schemes where one of the two parameters is kept fixed.
This highlights the danger of neglecting neutrino mass.

In each of the three scenarios analysed here, we shall study the
impact of each data set separately, devoting special attention to parameter degeneracies
and to the impact of galaxy clustering data.
We shall also study the preferred values
of $\neff$ arising from our different numerical studies.
As stated above, $\neff$ has been claimed to be larger than the standard model expectation in a plethora of cosmological data analyses.
We shall see that using up-to-date cosmological data, the evidence for $\neff>3$ still remains at the $2\sigma$ level.
The data sets exploited include recent measurements of the BAO feature in the Data Release 9 (DR9; \cite{DR9}) CMASS sample of the Baryon Oscillation Spectroscopic Survey (BOSS)~\cite{boss, boss2012}, which is part of SDSS-III~\cite{sdss3} with a median redshift of $z=0.57$~\cite{anderson}.
 
The structure of the paper is as follows. Section~\ref{sec:model} describes the three standard and non-standard physics scenarios considered here, summarizing  the main effects of the neutrino parameters on the cosmological observables. 
In Sec.~\ref{sec:data} we describe the data sets used in the numerical analyses and in Sec.~\ref{sec:i} the constraints on the neutrino thermal abundance $\neff$ and on the sum of their masses are presented for different combinations of data in the three schemes studied here. We draw our conclusions in Sec.~\ref{sec:concl}.

\section{The Model: Neutrinos in Cosmology}
\label{sec:model}
We shall start with the standard scenario, which is made of three massive neutrinos and put constraints on the sum of their masses $\sum m_\nu$. We therefore fix $\neff=3.04$. Current neutrino oscillation data can not distinguish between the two possible types of ordering of the neutrino mass eigenstates, i.e.~information
regarding the neutrino mass hierarchy is inaccessible. While the effect of neutrino mass on the CMB is related to the total amount
of energy density in the form of massive neutrinos and the individual values of the neutrino masses are irrelevant,
the matter power spectrum in principle provides information on the sum of the neutrino masses as well as on
the individual values of the $m_{\nu_{i}}$'s. The non-relativistic neutrino overdensities cluster
at a given redshift $z$ only at scales where the wavenumber of perturbations is below the neutrino free streaming scale
\begin{equation}
k_{f_s}(z)=\frac{0.677}{(1+z)^{1/2}}\left(\frac{m_\nu}{1\ \textrm{eV}}\right)
\sqrt{\Omega_{m}} h \ \textrm{Mpc}^{-1}~,
\end{equation}
due to the pressure gradient, which prevents gravitational clustering ($\Omega_m$ is the ratio of the total matter energy density over the critical density at redshift zero). On spatial scales larger than the free streaming scale $k < k_{f_s}$, neutrinos cluster and behave as pressureless matter (dust). Perturbations with comoving wavenumber larger than the free streaming scale can not grow due to the large neutrino velocity dispersion. As a consequence, the growth rate of density perturbations decreases and the matter power spectrum is suppressed at $k>k_{f_s}$. Therefore, there is a clear signal on the matter power spectrum induced by the presence of neutrino masses. Neutrinos of different masses will have different transition redshifts from relativistic to non-relativistic behavior, and, in principle, by exploiting the information contained on the matter power spectrum it could be possible to identify the neutrino mass hierarchy and isolate the individual neutrino masses. However, in practice, such a task has been shown to be extremely challenging~\cite{Jimenez:2010ev}. In the following, we will assume a degenerate mass spectrum. In the standard scenario with three active massive neutrinos, the parameters considered in the analysis are:

\begin{equation}
 \label{parameter}
\{\omega_b,\omega_c, \Theta_s, \tau, n_s, \log[10^{10}A_{s}], \sum m_\nu\}~,
\end{equation}
where $\omega_b\equiv\Omega_bh^{2}$ and $\omega_c\equiv\Omega_ch^{2}$ are the physical baryon and cold dark matter densities,
$\Theta_{s}$ is the ratio between the sound horizon and the angular diameter distance at decoupling, $\tau$ is the optical depth,
$n_s$ is the scalar spectral index, $A_{s}$ is the amplitude of the primordial spectrum and $\sum m_{\nu}$ is the sum of the active neutrino masses.

Next, we explore non-standard neutrino scenarios in which the effective number of thermalized species is parameterized by $\neff$. 
As commonly assumed in the literature, we start here assuming a massless neutrino scenario with $\neff$ massless species. One of the main effects of $\neff$ on the CMB temperature anisotropies arises from the change of the epoch of the radiation matter equality, shifting therefore the location of the CMB acoustic peaks. This position is given by the so-called acoustic scale $\theta_A$, which reads
\begin{equation}
\theta_A=\frac{r_s(z_{rec})}{r_\theta(z_{rec})}~,
\end{equation}
where $r_\theta (z_{rec})$ and $r_s(z_{rec})$ are the comoving angular diameter distance to the last scattering surface and the sound horizon at the recombination epoch $z_{rec}$, respectively. Although $r_\theta (z_{rec})$ almost remains the same for different values of $\neff$, $r_s(z_{rec})$ becomes smaller when $\neff$ is increased. Therefore, the positions of acoustic peaks are shifted to higher multipoles (smaller angular scales) if the value of $\neff$ is increased. However, this  effect can be compensated by changing the cold dark matter density, in such a way that $z_{rec}$ remains fixed, see Ref.~\cite{Hou:2011ec}. Therefore, due to the degeneracy with the cold dark matter component, the change induced at low $\ell$ is negligible and the largest impact of $\neff$ on the CMB temperature anisotropies comes from its effect on high multipoles $\ell$, since a higher value of $\neff$ will induce a drop in power at small scales due to an increased Silk damping. Silk damping refers to the suppression in power of the CMB temperature anisotropies on scales smaller than the photon diffusion length. As a random walk process, the diffusion distance $r_d$ will increase as the square root of time and therefore a higher expansion rate $H(z)$ (caused by a higher $\neff$) will \emph{decrease} $r_d$, decreasing therefore the damping. However, in order to keep the acoustic scale $\theta_A$ fixed, the comoving angular $r_\theta (z_{rec})$ distance at decoupling should scale in the same way as $r_s(z_{rec})$ does (i.e.~as $1/H(z)$) and the angular scale of the Silk damping $\theta_d= r_d/r_\theta (z_{rec})$ will grow as $\sqrt{H}$. Consequently, the damping will be \emph{increased} when $\neff$ does due to the higher expansion rate $H(z)$, see Ref.~\cite{Hou:2011ec}.    

This first non-standard scenario is described by the following set of parameters:

\begin{equation}
 \label{parameter}
 \{\omega_b,\omega_c, \Theta_s, \tau, n_s, \log[10^{10}A_{s}], \neff\}~.
\end{equation}

However, we know from neutrino oscillation experiments that neutrino have non-zero masses and mixings. In fact, a recent neutrino oscillation analysis has shown that models with three active plus two sterile light (sub-eV) neutrino species provide a very good fit to short baseline data~\cite{Kopp:2011qd}. Therefore the previous minimal non-standard scenario with $\neff$ massless neutrinos should be enlarged to accommodate active plus sterile neutrino masses and mixings. The last scenario explored here therefore
consists of $\neff$ massive neutrino species, each with equal\footnote{Also in the presence
of additional species, current data are not sensitive to how the masses are distributed over the species, but only
to the sum of the masses and the number of species. This justifies the use of a degenerate mass spectrum.} mass $m_\nu$.
The effects of these two parameters on the observables have been summarized above. While the highest sensitivity to
the neutrino masses comes from large scale structure information, and the highest $\neff$ sensitivity comes from CMB measurements,
there will be significant covariance between the two parameters
and we explore this degeneracy in \ref{subsubsec:b}. In this last non-standard scenario, the set of parameters involved in the study are

\begin{equation}
 \label{parameter}
  \{\omega_b,\omega_c, \Theta_s, \tau, n_s, \log[10^{10}A_{s}], \neff, \sum m_\nu\}~.
\end{equation}
For our numerical analyses, we have used the Boltzmann CAMB code~\cite{camb} and extracted cosmological parameters from current data
using a Monte Carlo Markov Chain (MCMC) analysis based on the publicly available MCMC package \texttt{cosmomc}~\cite{Lewis:2002ah}.
Table \ref{tab:priors} specifies the priors considered on the different cosmological
parameters~\footnote{In the next section we will describe the additional nuisance parameters that we have also added in the analyses.
These parameters are related to some cosmological data sets exploited here.}.
Our neutrino mass prior is cast in the form of a (uniform) prior on the neutrino density fraction $f_{\nu} = \Omega_{\nu}/\Omega_{\rm DM}$,
where $\Omega_{\nu}$ is the ratio of the neutrino energy density over the critical density at redshift zero, and $\Omega_{\rm DM}$
is the same ratio, but for the total dark matter density, which includes cold dark matter and neutrinos.

We restrict ourselves to a $\Lambda$CDM cosmology in the different
scenarios described above. 

\begin{table}[h!]
\begin{center}
\begin{tabular}{c|c}
\hline\hline
 Parameter & Prior\\
\hline
$\Omega_{b}h^2$ & $0.005 \to 0.1$\\
$\Omega_{c}h^2$ & $0.01 \to 0.99$\\
$\Theta_s$ & $0.5 \to 10$\\
$\tau$ & $0.01 \to 0.8$\\
$n_{s}$ & $0.5 \to 1.5$\\
$\ln{(10^{10} A_{s})}$ & $2.7 \to 4$\\
$f_\nu$ &  $0 \to 0.2$\\
$\neff$ &  $1.047 \to 10$\\
\hline\hline
\end{tabular}
\caption{Uniform priors for the cosmological parameters considered here.}
\label{tab:priors}
\end{center}
\end{table}

\section{Data}
\label{sec:data}

Our baseline data set is the seven--year WMAP data \cite{Komatsu:2010fb,wmap7}  (temperature and polarization)
with the routine for computing the likelihood supplied by the WMAP team. We then also add CMB data from the
South Pole Telescope (SPT)~\cite{spt}, which strongly improve the measurement of the temperature anisotropies on
scales $\lesssim 10$ arcmin. In order to address foreground contributions, the SZ amplitude $A_{SZ}$,
the amplitude of the clustered point source contribution, $A_C$, and the amplitude of the Poisson distributed point source contribution, $A_P$, are added as nuisance parameters in the CMB data analyses. We have followed Ref.~\cite{spt}, applying a Gaussian prior on the amplitude of each of these three foreground terms. 

To the WMAP basic data set we add the latest constraint on the Hubble constant $H_0$ from the Hubble Space Telescope (HST)~\cite{Riess:2011yx}, and supernova data from
the 3 year Supernova Legacy Survey (SNLS3), see Ref.~\cite{snls3}. In the case of SNLS3 data, we add in the MCMC analysis two
extra nuisance parameters related to the light curve fitting procedure used to analyze the supernova (SN) data.
These parameters characterize the dependence of the intrinsic supernova magnitude on stretch (which measures the shape of the SN light curve) and color~\cite{snls3}.

We also consider data from galaxy clustering that we summarize next.
We first employ data from SDSS-II  (Sloan Digital Sky Survey; \cite{yorketal2000}) Data Release 7 (DR7; \cite{DR7}),
analyzing separately the full shape of the halo power spectrum derived from the clustering of luminous red galaxies~\cite{beth}, and the
Baryon Acoustic Oscillation (BAO) signal as extracted from the same data set~\cite{percival}.
Then we consider the impact of the most recent measurement of the BAO scale \cite{anderson} from the CMASS sample in Data Release 9 (DR9; \cite{DR9})
of the Baryon Oscillation Spectroscopic Survey (BOSS)~\cite{boss, boss2012}, with a median
redshift of $z=0.57$. Together with the CMASS DR9 data, we also include the recent measurement of the BAO scale based on a re-analysis (using reconstruction \cite{eisetal07})
of the LRG sample from Data Release 7 with a median redshift of $z=0.35$~\cite{nikhil}, and the measurement of the BAO signal at a lower
redshift $z=0.106$ from the 6dF Galaxy Survey 6dFGS~\cite{6dFGS}. We will refer to the combination of these three new BAO signals as BAO$_{2012}$.

\section{Results}
\label{sec:i}
Here we present the constraints from current data on the neutrino thermal abundance $\neff$ and on the sum of their masses in different scenarios.

\subsection{The standard lore: Three massive neutrinos}

In this section we consider the standard picture, i.e.~three massive neutrinos
and put constraints on the sum of their masses $\sum m_\nu$. 

\begin{figure}
\includegraphics[width=9.cm]{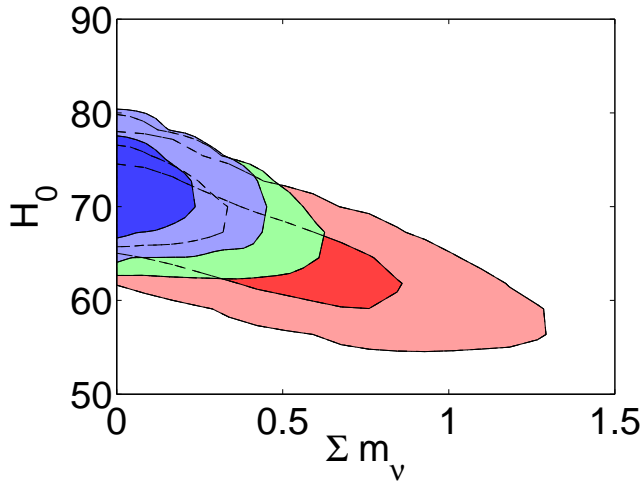}
\caption{The red contours show the 68\%  and 95\% CL constraints in the ($\sum m_\nu$, $H_0$) plane from our basic WMAP data set. The blue contours show the results from the combination of WMAP and HST data, while the green contours depict the results from the combination of WMAP and SNLS3 data sets. Notice that the strong degeneracy present in the case of WMAP data alone gets alleviated when a prior on H$_0$ from HST data is added in the analysis. SNIa luminosity distante data are unable to independently determine the Hubble constant $H_0$ but measure the $\Omega_{m}$ quantity, and, since WMAP measures $\Omega_{m} h^2$, the combination of WMAP plus SNIa data is able to determine $H_0$.}
\label{fig:fig0}
\end{figure}

Table~\ref{tab:constraints_standard_1} shows the $68\%$ and $95\%$ CL bounds on 
$\sum m_\nu$ from several combinations of WMAP data with SPT, HST and SNLS3.
These constraints agree quite well with previous analyses in the literature~\cite{dePutter:2012sh}.
Note that WMAP alone sets a $95\%$ CL bound of $\sim 1$~eV on the sum of neutrino masses.
When HST data is added in the analysis the results improve in a very significant way, since HST data helps
enormously in breaking the strong degeneracy between $\sum m_\nu$ and $H_0$, see Fig.~\ref{fig:fig0}.
Indeed, without a $H_0$ measurement, the change induced in the CMB temperature anisotropies caused by an increase in $\sum m_\nu$ can be
compensated by a decrease in $H_0$. The reason is the following:
if $\sum m_\nu$ increases, the main effect on the CMB comes from the resulting
shift in the distance to last scattering\footnote{$r_{\theta}(z_{\rm rec}) \propto \int_0^{z_{\rm rec}} dz \, \left[\omega_r a^{-4} + \omega_m a^{-3} + (1 - \omega_m/h^2) \right]^{-1/2}$,
with $\omega_m = \omega_b + \omega_c + \omega_\nu$}.
While the acoustic peak structure of the CMB data does not leave much freedom in $\omega_c$ and $\omega_b$,
the change in distance can be compensated by lowering $h$ (i.e.~$H_0$). The HST prior on the Hubble parameter will break this strong degeneracy,
setting a $\sim 0.4$~eV 95\% CL bound on the sum of the three active neutrino masses.
A similar result but to a lesser extent arises when combining WMAP and SNLS3 data sets. Note that we follow here a conservative approach avoiding the combination of HST and SNLS3 data, since these two data sets are not totally independent. The addition of SPT data does not help in constraining $\sum m_\nu$: taking into account the marginalization over the additional nuisance parameters, the high $\ell$ multipole data are not sensitive to neutrino mass, nor do they help break degeneracies to improve the neutrino bound and therefore we will not consider SPT data set in the rest of the analyses presented in this section. Of course this will not be the case when the effective number of neutrino species is added as a free parameter in the analyses, since, in that case, SPT high $\ell$ measurements will become crucial, as the combination of WMAP and HST will not be enough to break the strong degeneracy between $\neff$ and $\sum m_\nu$.

Table \ref{tab:constraints_standard_2} shows the constraints in the standard massive neutrino scenario when SDSS-II galaxy clustering data are added in the analysis. Note that the bounds from \emph{geometrical} information from the BAO signal are very similar to those obtained from the \emph{\emph{shape}} of the galaxy power spectrum measurements, although BAO information provides slightly better constraints on the sum of the three active neutrino masses. These findings agree with previous analyses in the literature, see e.g. Ref.~\cite{Hamann:2010pw}. However, we are considering here a minimal cosmological model, with spatial flatness and a cosmological constant. BAO/\emph{geometrical} data will not be as powerful as \emph{shape} measurements when other cosmological parameters are added in the analysis, since the information contained in the full \emph{shape} of the matter power spectrum is extremely useful for breaking degeneracies. We will illustrate this effect in section \ref{subsubsec:b}, where the addition of the number of massive neutrino species $\neff$ will open up new degeneracies, and the matter power spectrum \emph{shape} information will be more powerful in resolving them than the \emph{geometrical} BAO measurements. Note from Tab.~\ref{tab:constraints_standard_2} that if other data sets are included, such as, for instance, a prior on $H_0$ from the HST experiment, some degeneracies are broken, and the constraints on $\sum m_\nu$ from \emph{shape} and \emph{geometrical} SDSS-II measurements of the galaxy distribution become very similar again.
In fact, once the HST prior is included, further inclusion of the large scale structure data considered here does not significantly improve the neutrino mass bound in the first place.

\begin{figure}
\includegraphics[width=9.cm]{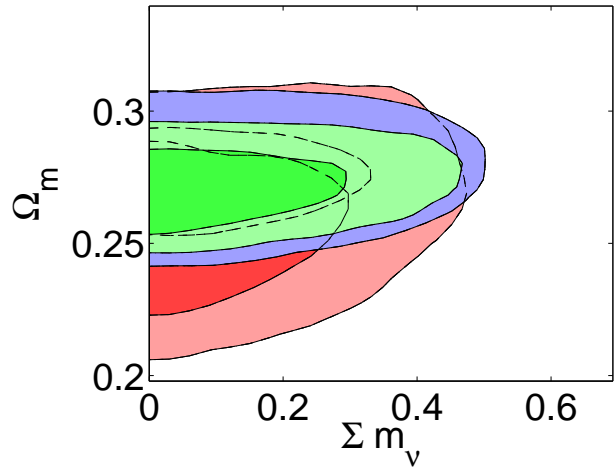}
\caption{The red contours show the 68\%  and 95\% CL constraints in the ($\sum m_\nu$, $\Omega_m$) plane from WMAP plus HST data sets. The blue contours show the results from the combination of WMAP, HST and galaxy clustering measurements from SDSS-II interpreted in the form of BAO signals, while the green contours depict the results from the combination of WMAP, HST and BAO$_{2012}$ data sets. The errors on the $\Omega_m$ parameter are significantly improved when considering either the SDSS-II BAO or the BAO$_{2012}$ measurements. Also, due to the higher mean value of $\Omega_m$ when adding BAO information to the WMAP and HST measurements, a slightly higher neutrino mass is allowed. }
\label{fig:fig0b}
\end{figure}

Table \ref{tab:constraints_standard_3} shows the constraints on $\sum m_\nu$ with the addition of the new BAO data from the CMASS DR9 BOSS experiment combined with a re-analysis of the LRGDR7 sample and the BAO signal from 6dFGS. Note that the bounds in this minimal cosmological model from the BAO$_{2012}$ data set combination are slightly better, albeit very close to those obtained with BAO SDSS-II measurements. Note from Tabs. \ref{tab:constraints_standard_2} and \ref{tab:constraints_standard_3} that the combination of WMAP and HST data with old BAO measurements  gives a bound on $\sum m_\nu$ which is worse than the one arising from the combination of WMAP plus HST alone. The errors on the $\Omega_m$ parameter are significantly improved when considering either the SDSS-II BAO or the BAO$_{2012}$ measurements, see Fig.~\ref{fig:fig0b}. The mean value of $\Omega_m$ is also higher when adding BAO information to the WMAP and HST measurements and, consequently, a slightly higher neutrino mass is allowed. 

\begin{table*}
\begin{center}
\begin{tabular}{lccccccc}
\hline \hline
   &            & WMAP & WMAP+HST & WMAP+SPT & WMAP+SNLS3 \\
\hline
\hspace{1mm}\\
$\sum m_\nu$ (eV) & 68\% CL & $<0.60 $ & $<0.16$ & $<0.52$ & $<0.24$ \\
                & 95\%  CL &  $< 1.05$ &$<0.37$ & $<1.07 $ & $<0.51$ \\ 
\hline
\hline
\end{tabular}
\caption{Constraints on ${\mnu}$ from WMAP data alone and WMAP combined with HST measurements of the Hubble constant, SPT data and SNLS3 data. We have fixed the number of massive neutrinos to 3.046.}
\label{tab:constraints_standard_1}
\end{center}
\end{table*}

\begin{table*}
\begin{center}\footnotesize
\begin{tabular}{lccccccc}
\hline \hline
 &   & WMAP+MPK & WMAP+BAO &WMAP+MPK+HST&WMAP+BAO+HST&WMAP+MPK+SNLS3&WMAP+BAO+SNLS3\\
\hline
\hspace{1mm}\\
               
$\sum m_\nu$ (eV) & 68\% CL& $<0.37$  & $<0.29$ &$<0.16$ &$<0.22$ & $<0.22$ & $<0.27$\\  
                & 95\% CL & $<0.63$ & $<0.52$ & $<0.32$&$<0.42$& $<0.42$ & $<0.47$\\  
\hline
\hline
\end{tabular}
\caption{Constraints on the sum neutrino masses $\sum m_\nu$ from WMAP data and different possible combinations of galaxy clustering measurements from SDSS-II (interpreted either in the form of matter power spectrum measurements (MPK) or in the form of BAO signals (BAO)), HST and SNLS3 data sets. We have fixed the number of massive neutrinos to 3.046.}
\label{tab:constraints_standard_2}
\end{center}
\end{table*}
\begin{table*}
\begin{center}\
\begin{tabular}{lcccc}
\hline \hline
   &           & WMAP+BAO$_{2012}$  & WMAP+BAO$_{2012}$+HST  & WMAP+BAO$_{2012}$+SNLS3 \\
\hline
\hspace{1mm}\\
               
$\sum m_\nu$ (eV) & 68\% CL& $<0.26$  & $<0.19$ & $<0.25$  \\
                & 95\% CL & $<0.48$ & $<0.38$ & $<0.47$ \\
\hspace{1mm}\\
\hline
\hline
\end{tabular}
\caption{Constraints on $\sum m_\nu$ from WMAP and different possible combinations of the new BAO data explored here (BAO$_{2012}$), HST and SNLS3 data sets. We have fixed the number of massive neutrinos to 3.046.}
\label{tab:constraints_standard_3}
\end{center}
\end{table*}
\subsection{Beyond the standard lore: varying $N_{\rm eff}$}

\subsubsection{Neglecting neutrino mass}

We assume here a scenario with $\neff$ massless neutrino species. Table~\ref{tab:constraints_massless_1} shows the central values
and the $68\%$ and $95\%$ CL errors on the quantity $\neff$ for our basic WMAP data set alone and also with other data sets explored here.
We observe that WMAP data alone is unable to constrain the value of $\neff$ but the situation drastically improves when adding either SPT or HST data. The reason for that is because $\neff$ is highly degenerate with both $H_0$ and the dark matter energy density $\Omega_{\rm DM} h^2$ at the multipole range relevant to WMAP CMB measurements: the change induced in the CMB spectrum by $\neff$ can be compensated by a change in the value of the Hubble constant $H_0$. HST data is crucial for breaking the strong degeneracies between $\neff$ and both the physical amount of matter $\Omega_{\rm DM} h^2$ and the Hubble constant $H_0$, therefore, the prior on $H_0$ from HST improves significantly the constraints on $\neff$. In the following section we will see when adding $\sum m_\nu$ as a free parameter the errors on $\neff$ from WMAP plus HST will be much higher, due to the strong degeneracy between the effective number of neutrino species and the sum of the neutrino masses.

SPT data is also very powerful in constraining $\neff$ due to the information contained at high multipoles $\ell$. As summarized in Sec.~\ref{sec:model}, an increasing value of $\neff$ has little (big) impact on low (high) multipoles $\ell$ due to an increased Silk damping caused by a higher expansion rate. Therefore, SPT high multipole data helps enormously in constraining the value of $\neff$. The combination of WMAP, SPT and HST gives $3.83^{+0.21 +0.86}_{-0.25 -0.76}$ which deviates $\sim 2 \sigma$ from the expected standard model value for $\neff$. 
 The combination of CMB measurements with SNLS3 luminosity distance data results in a mean value of $\neff$ which is only $\sim 1 \sigma$ away from its the standard value. 

The addition of the information contained in the matter power spectrum or in the BAO signal from SDSS-II galaxy clustering measurements to our basic WMAP data set does not change
significantly the central values nor the errors depicted in Tab.~\ref{tab:constraints_massless_1} and  for the sake of clarity we do not show all the possible combinations here.
We only illustrate the case with the addition of the new BAO data from the BOSS DR9 CMASS sample combined with a re-analysis of the LRGDR7
sample and the BAO signal from 6dFGS, see Tab.~\ref{tab:constraints_massless_3}. The evidence for extra radiation $\neff>3$ claimed in the literature still remains at the $2\sigma$ level, in perfect agreement with a recent analysis presented in Ref.~\cite{latest}.  
 
\begin{table*}
\begin{center}
\begin{tabular}{lcccccc}
\hline \hline
   &            & WMAP &WMAP+HST & WMAP+SPT & WMAP+SPT+HST & WMAP+SPT+SNLS3  \\
\hline
\hspace{1mm}\\
${\neff}$ &   & $5.99^{+4.01 +4.01}_{-1.23 -3.47}$ & $4.19^{+0.33 +1.30}_{-0.36 -1.26}$ & $4.13^{+0.35 +1.46}_{-0.41 -1.30}$ & $3.83^{+0.21 +0.86}_{-0.25 -0.76}$  & $3.51^{+0.31 +1.25}_{-0.35 -1.16}$\\
\hspace{1mm}\\
\hline
\hline
\end{tabular}
\caption{Constraints on ${\neff}$ for the massless neutrino scenario from WMAP data, WMAP data and HST measurements of the Hubble constant, WMAP and SPT data and WMAP plus SPT plus SNLS3 data.}
\label{tab:constraints_massless_1}
\end{center}
\end{table*}

\begin{table*}
\begin{center}\
\begin{tabular}{lccccc}
\hline \hline
            &           & WMAP+BAO$_{2012}$  & WMAP+SPT   & WMAP+SPT  & WMAP+SPT \\
            &            &                & +BAO$_{2012}$                & +BAO$_{2012}$+HST   & +BAO$_{2012}$+SNLS3\\
\hline
\hspace{1mm}\\
               
${\neff}$ &  & $5.86^{+1.07 +3.51}_{-1.15 -3.36}$ & $3.48^{+0.31 +1.27}_{-0.36 -1.17}$ & $3.66^{+0.20 +0.73}_{-0.21 -0.69}$  &  $3.52^{+0.31 +1.30}_{-0.38 -1.16}$\\
\hspace{1mm}\\
\hline
\hline
\end{tabular}
\caption{Constraints on $\neff$ from WMAP and different possible combinations of the new BAO data explored here (BAO$_{2012}$), HST and SNLS3 data sets.}
\label{tab:constraints_massless_3}
\end{center}
\end{table*}

\subsubsection{Including neutrino mass}
\label{subsubsec:b}
We consider here a scenario with $\neff$ massive species, each with a mass $m_\nu$. Table \ref{tab:constraints_massive_1} shows the joint constraints on the sum of neutrino masses $\sum m_\nu$ and on the number of massive neutrino species $\neff$ when HST, SPT and SNIa data are added to our basic WMAP data set. Note that the addition of HST data to the WMAP basic data set does not help as in the massless neutrino scenario, the reason for that being that $m_\nu$ is now a free parameter and it is strongly degenerate with $H_0$ (among other parameters, such as the matter energy density). Therefore, a prior on $H_0$ is not enough to improve the constraints from WMAP alone. The errors on $\neff$ are an order of magnitude larger than in the massless neutrino scenario. A similar effect occurs when addressing the neutrino mass bound: in the case where only three active massive neutrinos were considered, the WMAP and HST data combination provided a 95\% CL bound $\sum m_\nu < 0.38$~eV. The former bound becomes $2.25$~eV when $\neff$ is allowed to vary as well.

High multipole SPT data helps in constraining $\neff$. SNIa data is very useful
with respect to neutrino masses. However,
both the mean value of $\neff$ and its error are very large (due to the strong degeneracy between $\neff$ and $\sum m_\nu$ as we shall see below)
unless SPT data is added in the analysis.
Such an addition (i.e.~WMAP+SPT+SNLS3) allows to partially break the strong degeneracy between $\sum m_\nu$ and $\neff$ and it
decreases both the mean value of $\neff$ and its errors significantly.
Note from Table \ref{tab:constraints_massive_1} that the neutrino mass bounds from CMB measurements plus SNIa data are better than those from CMB plus HST measurements. After the addition of SNLS3 measurements, the data combination prefers a lower value of $\Omega_m$ which, in turn, leads to a smaller value of $\sum m_\nu$.

Table \ref{tab:constraints_massive_2} shows the constraints in the massive neutrino scenario when data from galaxy clustering measurements
are added in the analysis. When considering SDSS-II measurements, the bounds are stronger (about a factor of two) if the information from galaxy clustering is
exploited as \emph{shape} constraints (i.e.~power spectrum measurements) rather than as \emph{geometrical} measurements (i.e.~BAO signals),
the reason for that being the presence of more degeneracies involved, as previously explained. When other data sets are included,
some degeneracies are broken, and the constraints from \emph{shape} and \emph{geometrical} SDSS-II measurements of the galaxy distribution
are very similar. We also depict the bounds on the sum of neutrino masses and on the number of massive species when new BAO information from
the BOSS DR9 CMASS sample, combined with a re-analysis of the LRGDR7 sample and with the BAO signal from 6dFGS, are added in the analysis.
Note that these new BAO data sets are more powerful than the previous BAO data set because they combines BAO measurements at several redshifts,
rely on a larger volume, and include an improvement of the original SDSS-II BAO measurement. Because of this, the BAO$_{2012}$ constraints are even better than those obtained with the full matter power spectrum of SDSS-II.

\begin{figure*}[h]
\begin{tabular}{c c}
\includegraphics[width=7.5cm]{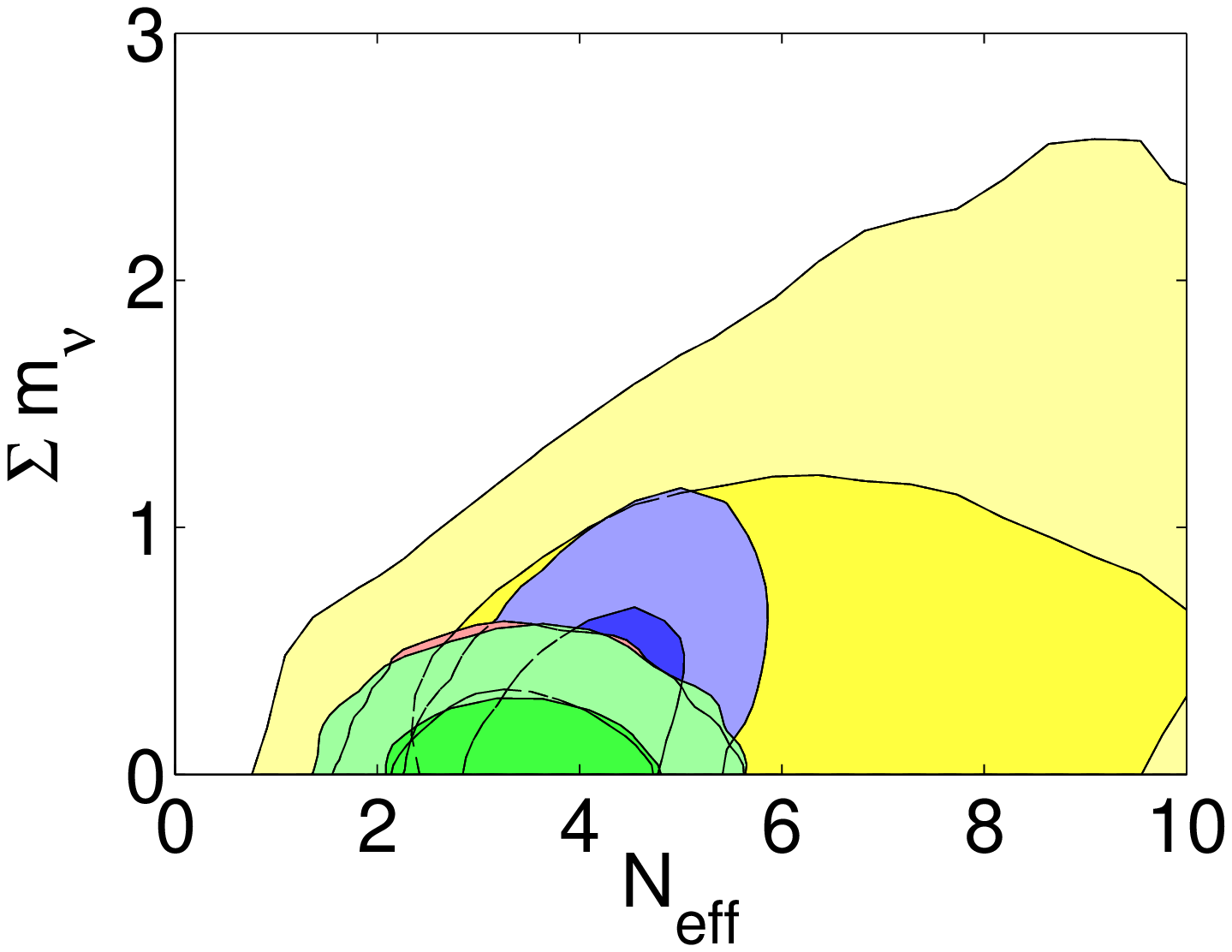}&
\includegraphics[width=7.5cm]{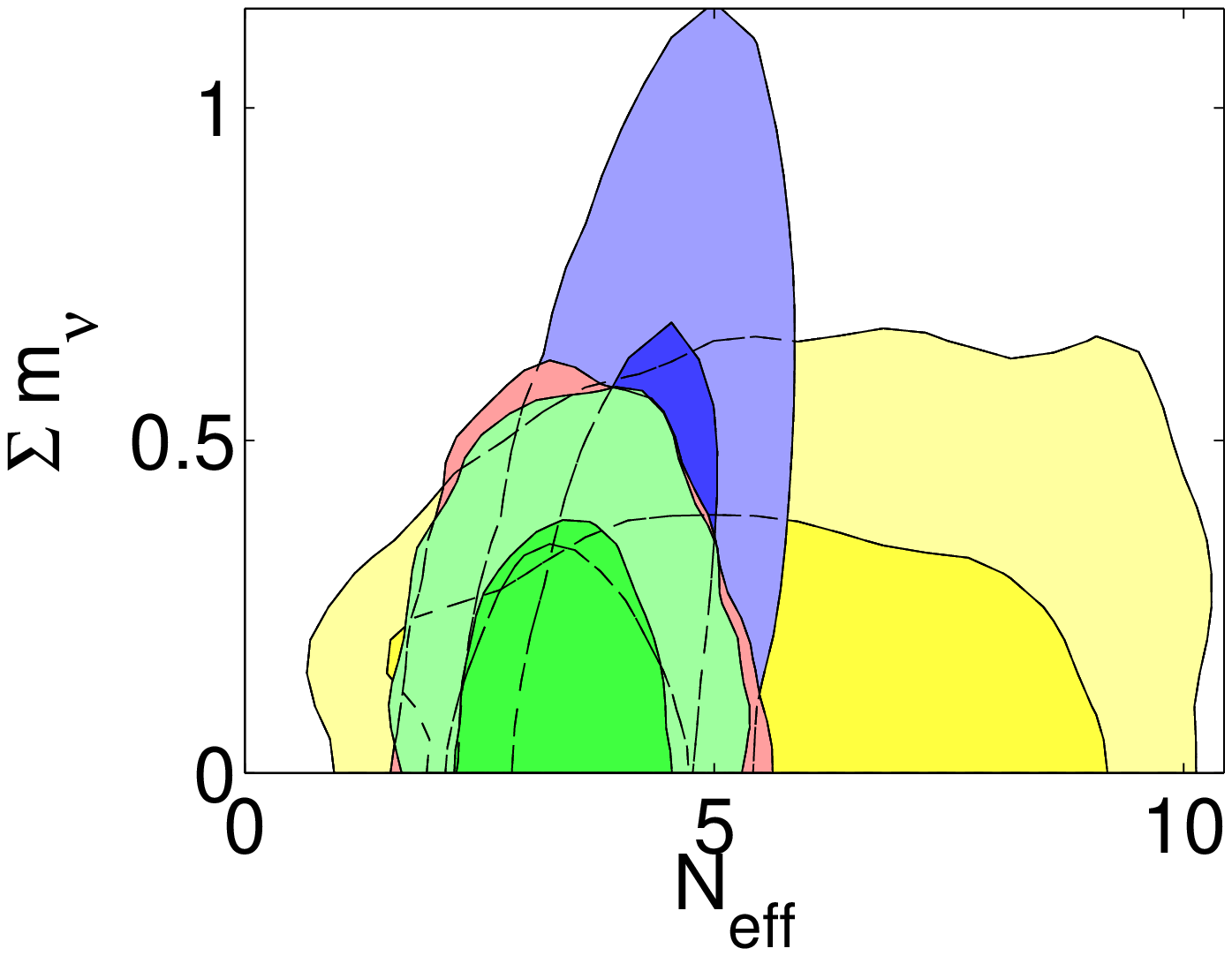}\\
\end{tabular}
\caption{Left panel: the yellow contours show the 68\%  and 95\% CL constraints in the ($\neff, \sum m_\nu$) plane from our basic WMAP data set combined with old BAO data. The blue contours show the results from the combination of WMAP, SPT and HST measurements, while the red contours depict the results from the combination of WMAP, SPT and SNLS3 measurements. Finally, the green contour denotes the constraints from the combination of WMAP, SPT, SNLS3 and BAO data. Right panel: as in the left panel but using recent BAO$_{2012}$ data.} 
\label{fig:deg}
\end{figure*}

The strong degeneracy between $\neff$ and $\sum m_\nu$ is illustrated in Fig.~\ref{fig:deg}, where we show the combination of WMAP with new and old BAO data sets as well as the combination of CMB data with HST or SNLS3 measurements. Finally, we also show the combination of CMB, BAO (both new and old data sets) and SNLS measurements.
As in the massless neutrino case, the addition of new BAO$_{2012}$ data to CMB and HST measurements makes the value of $\neff$ (marginally) consistent
with the standard expectation within $2 \sigma$, in agreement with previous analyses in the literature. Therefore the evidence for $\neff>3$ still persists when considering massive neutrino species. The 95\% CL constraints on the effective number of massive neutrino species and on the sum of their masses from CMB measurements, SNLS3 and BAO$_{2012}$ data are  $3.44^{+1.20}_{-1.24}$ and $\sum m_\nu<0.47$~eV, respectively. The former bounds translate into  $3.71^{+0.74}_{-0.72}$ and $\sum m_\nu<0.51$~eV when CMB, HST and BAO$_{2012}$ measurements are considered.  The bounds and the errors found in this work on both $\sum m_\nu$ and $\neff$ are similar to those presented in a recent study presented in Ref.~\cite{latest}, where a preference for $\neff >3$ at a 2$\sigma$ level has been presented. 

\begin{table*}
\begin{center}
\begin{tabular}{lccccccc}
\hline \hline
   &            & WMAP & WMAP+HST & WMAP+SPT & WMAP+SPT+HST & WMAP+SNLS3 & WMAP+SPT+SNLS3  \\
\hline
\hspace{1mm}\\
${\neff}$ &   & $6.18^{+3.82 +3.82}_{-1.18 -3.34}$ & $5.91^{+4.09 +4.09}_{-0.94 -2.44}$ & $3.91^{+0.30 +1.32}_{-0.39 -1.15}$ & $4.19^{+0.25 +1.02}_{-0.30 -0.92}$ & $5.97^{+4.03 +4.03}_{-1.17 -3.28}$ & $3.46^{+0.32 +1.31}_{-0.37 -1.17}$\\
\hspace{1mm}\\
${\sum m_\nu}$ (eV)& 68\% CL & $< 0.90$ & $<1.07$ & $< 0.81$ & $< 0.46$ & $< 0.28$ & $< 0.22$  \\
                & 95\%  CL &  $< 2.02$ &$<2.25$ & $< 1.47$ & $< 0.90$ & $< 0.66$ & $< 0.50$\\ 
\hline
\hline
\end{tabular}
\caption{Constraints on the number of massive neutrino species and on the sum of their masses $\sum m_\nu$ from WMAP data alone, WMAP data and HST measurements of the Hubble constant, WMAP and SPT data, WMAP and SNLS3 data, and WMAP plus SPT plus SNLS3 data.}
\label{tab:constraints_massive_1}
\end{center}
\end{table*}

\begin{table*}
\begin{center}\footnotesize
\begin{tabular}{lccccccccc}
\hline \hline
&            & WMAP+MPK & WMAP+BAO & WMAP+BAO$_{2012}$& WMAP+MPK & WMAP+ BAO &WMAP+BAO$_{2012}$& WMAP+BAO$_{2012}$\\
&   & & & & +SPT+SNLS3 &SPT+SNLS3&+SPT+HST&+SPT+SNLS3\\
\hline
\hspace{1mm}\\
               
${\neff}$ &  & $4.89^{+0.68 +3.18}_{-0.95 -2.42}$ & $6.18^{+3.82 +3.82}_{-1.15 -3.53}$ & $5.60^{+1.14 +3.68}_{-1.24 -3.55}$ & $3.63^{+0.29 +1.14}_{-0.32 -1.05}$ &$3.44^{+0.33 +1.24}_{-0.37 -1.14}$&$3.71^{+0.20 +0.75}_{-0.22 -0.72}$ & $3.44^{+0.33 +1.20}_{-0.37 -1.24} $\\
\hspace{1mm}\\
${\sum m_\nu}$ (eV) & 68\% CL& $<0.51$  & $<0.93$  & $<0.29$  & $<0.27$& $<0.21$& $<0.26$&$<0.25$\\
                & 95\% CL & $<1.07$ & $<2.11$  & $<0.59$ & $<0.50$& $<0.47$&$<0.51$ & $<0.47$\\
\hline
\hline
\end{tabular}
\caption{Constraints on the number of massive neutrino species and on the sum of their masses $\sum m_\nu$ from WMAP data and different possible combinations of galaxy clustering measurements from SDSS-II (interpreted either in the form of matter power spectrum measurements (MPK) or in the form of BAO signals (BAO)), HST and SNLS3 data sets.}
\label{tab:constraints_massive_2}
\end{center}
\end{table*}

\section{Conclusions}
\label{sec:concl}
We have presented constraints on several neutrino-extra radiation scenarios with the most recent and available cosmological data.
The data sets explored here include the recent measurement of the BAO feature in the DR9 CMASS sample of the BOSS experiment.
Three scenarios have been carefully examined: one standard scenario with three massive neutrino species, and two non-standard scenarios with
$\neff$ massless/massive neutrino species. In each scenario we have carefully explored the different existing degeneracies and the impact of the
different data sets exploited here, both separately and combined. In the standard scenario where only three active neutrinos are considered,
the combination of CMB WMAP plus HST data helps enormously in breaking the strong degeneracy between
$\sum m_\nu$ and the Hubble parameter $H_0$ that exists with CMB data only,
and the addition of further cosmological data sets does not improve significantly the limit on $\sum m_\nu$.
In this standard scenario, WMAP data plus geometrical information from BAO signals gives similar results to those obtained from
WMAP data plus the measurements
of the full shape of the matter power spectrum.  

In the non-standard scenario with $\neff$ massless neutrinos, the combination of WMAP and HST data is also able to break
the strong degeneracies between the effective number of neutrino species and other cosmological parameters, such
as the Hubble parameter and the physical dark matter energy density.
SPT high multipole CMB data is also extremely useful in constraining the effective number of neutrino species,
due to the increase in the Silk damping effect at small scales induced in the case of $\neff>3$.
The addition of SDSS-II galaxy clustering data or recent BAO measurements does not improve the constraints on $\neff$ in the massless neutrino scenario. 
We obtain $\neff=3.66^{+0.20 +0.73}_{-0.21 -0.69}$ when combining the former BAO data set with CMB and HST measurements. Therefore, the cosmological evidence for $\neff>3$ claimed often in the literature gets still remains when considering these new data sets, and the value of $\neff$ very close to the
standard model expectations ($\neff = 3.04$), being the discrepancy at the $2\sigma$ level.

In the most general non-standard scenario explored here, the strong degeneracy existing among the two neutrino parameters $\sum m_\nu$
and $\neff$ makes mandatory the combination of several data sets. The constraints from the combination of WMAP and HST measurements become
one order of magnitude weaker than in the previous two scenarios ($\neff$ fixed or $m_\nu$ fixed), although if more data sets are added in the analysis, the situation
improves significantly. Given that we know the sum of neutrino masses is non-zero, this means that
constraining the number of neutrino species while neglecting the sum of their masses
could lead to completely wrong results, and thus in general caution is
advised. In the reverse scenario, i.e.~constraining $m_\nu$ while keeping $N_{\rm eff}$ fixed, the same danger exists,
but the major difference is that $\neff$ may well be equal to its standard model value, while we know $\Sigma m_\nu$ is
non-zero.

To summarize the result with our maximal combination of data sets, the joint constraints on the effective number of massive neutrino species and on the sum of their masses from CMB, supernovae and the new BAO data are $3.71^{+0.75}_{-0.72}$ and $\sum m_\nu<0.51$~eV, respectively (both at the 95\% CL). Near future cosmological data, as CMB measurements from the Planck mission, combined with other probes, are expected to settle the issue of whether there exists cosmological compelling evidence for extra radiation.

\section{Acknowledgments}
We gratefully acknowledge Antonio Cuesta for providing the modified version of cosmomc with the recent BAO mesurements included.
We also thank Signe Riemer-S\o rensen and Chris Blake for their help
with the cross comparison of results.
O.M. is supported by the Consolider Ingenio project CSD2007-00060, by PROMETEO/2009/116, by the Spanish Ministry Science project FPA2011-29678 and by the ITN Invisibles PITN-GA-2011-289442.
Part of the research described in this paper was carried out at the Jet Propulsion Laboratory, California Institute of Technology,
under a contract with the National Aeronautics and Space Administration.



\end{document}